# Coupled Airy breathers

R. Driben[1,2,*], V. V. Konotop[3], and T. Meier[1]

[1] Department of Physics & CeOPP, University of Paderborn, Warburger Str. 100, D-33098 Paderborn, Germany
[2] ITMO University, 49 Kronverskii Ave., St. Petersburg 197101, Russian Federation
[3] Centro de Fisica Teorica e Computacional, Faculdade de Ciências, Universidade de Lisboa, Avenida Professor Gama Pinto 2, Lisboa 1649-003, Portugal
and Departamento de Fisica, Faculdade de Ciências, Universidade de Lisboa, Campo Grande, Lisboa, 1749-016, Portugal
*Corresponding author: driben@mail.uni-paderborn.de

The dynamics of two component coupled Airy beams is investigated. In the linear propagation regime a complete analytic solution describes breather like propagation of the two components featuring non-diffracting self-accelerating Airy behavior. The superposition of two beams with different input properties opens the possibility to design more complex non-diffracting propagation scenarios. In the strongly nonlinear regime the dynamics remains qualitatively robust as is revealed by direct numerical simulations. Due to the Kerr effect the two beams emit solitonic breathers, whose coupling period is compatible with the remaining Airy-like beams. The results of this study are relevant for the description of photonic and plasmonic beams propagating in coupled planar waveguides as well as for birefrigent or multi-wavelengths beams.

The original discovery of Airy waves, in the context of solutions to the linear Schrödinger equation [1], inspired extensive research for self-accelerating waves in optical settings [2-5]. Fascinating self-accelerating light beams propagating along the bending trajectories can manifest themself in the spatial and the temporal domains and are promising for a large variety of potential applications [6-9]. Recently, the observation of electron Airy beam has also been reported [10] as well as Airy waves in plasmonic structures [11, 12]. Plasmonic waves are usually confined to the metal-dielectric boundary. Plasmonic Airy beams are therefore characterized by a single transverse coordinate and a single propagation coordinate. Extending the concept of accelerating wavepackets into several dimensions has led to reports of the generation of linear light bullets [2, 13-16]. Under the strong influence of the Kerr nonlinearity in the high power propagation regime the Airy wave structure gets distorted, while the self-healing properties of Airy beams still demonstrate a strong resistance to the complete destruction [17]. Furthermore, solitons [18-23] and multi-solitons [24] shedding out of Airy pulses were reported while the front of the Airy pulse continued its propagation along its bended trajectory.

The objective of the present work is to demonstrate the dynamics of two component coupled Airy beams in linear and nonlinear media. In the linear propagation regime a complete analytic solution is presented which demonstrates periodically oscillating non-diffracting self-accelerating Airy beams – *Airy breathers*. The superposition of two beams with different input properties allows one to engineer peculiar complex robust beam evolution scenarios. In the strongly nonlinear regime it is demonstrated by direct numerical simulations that the dynamics remains qualitatively unchanged. Moreover, it is shown that, the two beams emit solitonic breathers, whose coupling period is compatible with the remaining Airy-like beams.

The dynamics of the coupled light waves with amplitudes, $u$ and $v$, propagating in $z$-direction in a third-order nonlinear media is described by coupled NLS equations which read in the normalized form:

$$iu_z = -u_{xx} + kv - 2g(|u|^2 + |v|^2)u \quad (1a)$$

$$iv_z = -v_{xx} + ku - 2g(|u|^2 + |v|^2)v \quad (1b)$$

where the parameter $k$ represents the linear coupling and the coefficient $g$ denotes the strength of the Kerr nonlinearity. For simplicity, we set $g = 1$, thus the influence of the Kerr effect will be driven solely by the amplitudes of the beams. Setting $k = 0$ the model becomes the well-known Manakov system [25] and is exactly integrable by means of the inverse scattering technique. Equations 1(a-b) with proper coefficients are relevant for the description of photonic and plasmonic beams propagating in coupled waveguides and for birefrigent or multi-wavelengths beams [26]. Equations 1 can describe the propagation of beams in two coupled planar dielectric waveguides, or the propagation of two coupled plasmonic beams in two parallel metal-dielectric boundary layers. Neglecting the nonlinearity ($g = 0$), i.e, restricting the analysis to the low-power linear regime, equations (1) are solved by:

$$u = 0.5\, exp(i\,k^2 z)\, (exp(ikz)\, U + exp(-ikz)\, V) \quad (2a)$$

$$v = 0.5\, exp(i\,k^2 z)\, (-exp(ikz)\, U + exp(-ikz)\, V) \quad (2b).$$

In the expressions above U and V denote arbitrary solutions of the linear Schrödinger equations (3):

$$iU_z = -U_{xx} \quad (3a)$$

$$iV_z = -V_{xx} \quad (3b)$$

We choose the following two solutions:

$$U = A * Airy(m(x - z^2))exp(-i\frac{2}{3}z^3 + ixz) \quad (4a)$$

$$V = A' * Airy(m'(x' - z^2))exp(-i\frac{2}{3}z^3 + ix'z) \quad (4b)$$

where $A$, $A'$ and $m$, $m'$ represent the amplitudes and the Airy argument parameters of the two beams, respectively. Also a displacement between the two beams can be considered with $x$ linearly shifted with a respect to $x'$.

We first consider the most fundamental case of propagation of both components in a low-power linear regime with equal beam parameters. Upper panels of Fig. 1 represent the beams dynamics with moderate coupling strength $k = 1$, while the lower panels display the case of stronger coupling with $k = 3$ and consequently show more frequent light transition between the components. We clearly observe that the light couples from one component to the other perfectly preserving its non-diffractive nature. Figs. 1-2 show the absolute values of the fields instead of intensities $|u|^2$, $|v|^2$ for better visibility of the peripheral areas of the beams. For Figs. 1-2 we assume that $|u|^2 \ll 1$ and $|v|^2 \ll 1$ and thus the nonlinear terms of equation (1) are negligible.

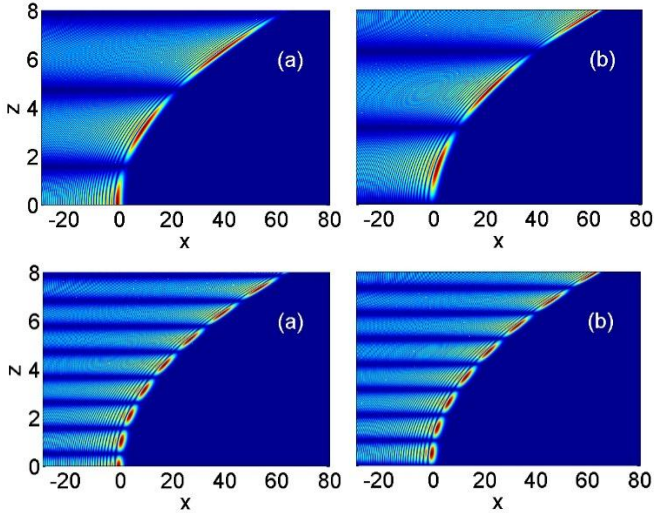

Fig. 1 (Color online). Evolution of Airy breathers of identical amplitude and initial position in the linear regime. The evolution of the u and v components is displayed in the (a) and (b) panels, respectively. The upper row shows the case of $m = 1$ and $k = 1$, while the lower row corresponds to $m = 1$ and $k = 3$.

Also solutions of equation (1) with different characteristics exist. The upper row of Fig. 2 reveals the dynamics for the case of an initial phase difference between the injected light in the two components such as $A' = \exp(i0.5\pi)A$, providing non-zero light power at the origin for both components. Making the amplitude of one of the component larger than that of the other will lead to partial light transfer as demonstrated in middle row of the Fig. 2 for $A' = 0.5A$. Taking one component's input spatially displaced relative to the other one along the x-axis results in more complex dynamical pattern with most intense peaks being not the leading ones as the lower row panels of the Fig. 2 demonstrates for $x' = x - 5$. One can clearly see that all cases considered here provide a diffraction free accelerating dynamics of both light beams.

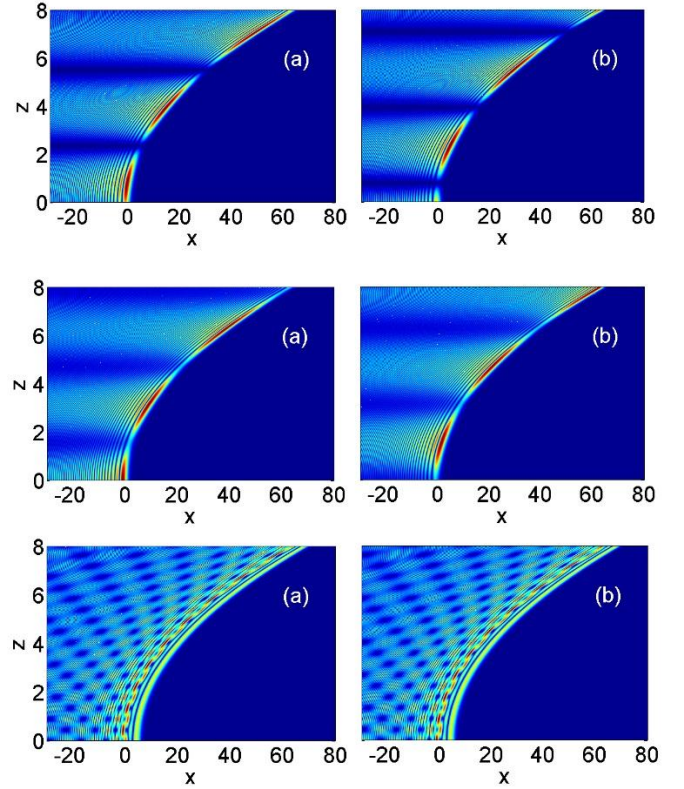

Fig. 2. (Color online) Evolution of Airy breathers in the linear regime with $m = 1$ and $k = 1$ and with manipulation of the initial parameters of the two components. The upper row shows the case with an input phase difference of $0.5\pi$. The middle row corresponds to the case with the amplitude of the v component equal to half of that of the u-component ($A' = 0.5A$). The lower row demonstrates the case of an initial lateral mismatch between the two components of $x' = x - 5$.

Analyzing the nonlinear propagation regime using the full eqs. (1) we observe robust breather like accelerating beams for a broad range of input amplitudes. At some stage ejections of solitonic breathers that arise from interactions between the lobes constructing the Airy beam start to appear [18-24]. This phenomenon can be viewed as an optical analog of Newton's cradle and was demonstrated in Ref. [27] to exist in other asymmetric light structures such as the evolution of dense multiple pulses with the same frequency and more importantly in fission of N-solitons under the dominant action of a third order dispersion. The upper panels of Fig. 3 display the dynamics of both components with $A = A' = 2$, while the lower panels correspond to a still stronger nonlinear regime with $A=A'=3$. Interestingly the ejected nonlinear waves share the same light power exchange mechanism with the same period as the main parts of the beams that continue their propagation. Taking into regards an initial lateral displacement between the two components of $x'=x-5$ in the strongly nonlinear regime leads to double vertically emitted breathers as the lower panels of Fig. 3 reveal.

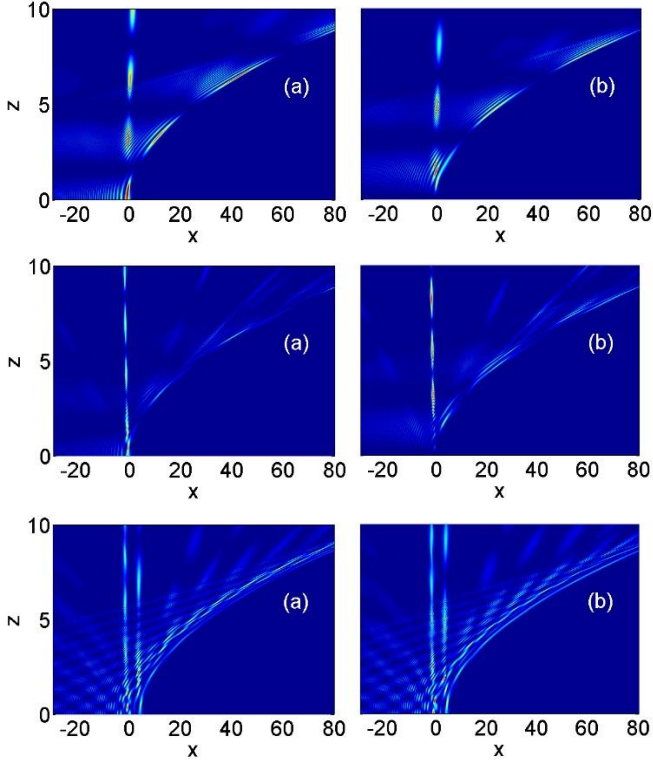

Fig. 3. (Color online) Evolution of Airy breathers in the nonlinear regime with $m=1$ and $k=1$. The upper row shows the case with $A=A'=2$. The middle row corresponds to $A=A'=3$ and the lower is also for $A=A'=3$ and includes an initial lateral mismatch between the two components of $x'=x-5$.

In conclusion, we have studied evolution of two component coupled Airy beams. In low-power, linear propagation regime we provide a full analytic description of non-diffractive breather-like dynamics of the two self-accelerating Airy beams. Manipulations by the input conditions of the two beams provide a possibility to create various peculiarly curved propagation trajectories for the two waves. In the high-power propagation regime the Kerr effect does not strongly perturb the robustness of the dynamics seen in the linear case. The two beams emit solitonic breathers, whose coupling period is compatible with the maintaining Airy-like beams. The results of this study are relevant for the description of photonic and plasmonic beams propagating in coupled waveguides and for the description of birefrigent or multi-wavelengths beams. Furthermore, our work paves a way to study other multicomponent self-accelerating beam dynamics such as Weber and Mathieu beams [28] and to analyze the evolution of multidimensional and/or multicomponent non-diffracting beams. Other related prospective routes of investigation are the dynamics of counter-propagating coupled non-diffracting beams and systems in which the coupling strength is modulated along the propagation direction.

The authors wish to express their gratitude to Ady Arie for the very fruitful discussions.
RD and TM gratefully acknowledge support provided by the Deutsche Forschungsgemeinschaft (DFG) via the Research Training Group (GRK 1464) and computing time provided by the PC² (Paderborn Center for Parallel Computing). RD gratefully acknowledges the support by the Government of the Russian Federation (Grant 074-U01) through ITMO Early Career Fellowship scheme.
VVK acknowledges support of the FCT ( Portugal ) grants PEst-OE/FIS/UI0618/2014.